\documentclass[twocolumn,showpacs,preprintnumbers,amsmath,amssymb,epsf]{revtex4-1}

\usepackage{graphicx,epsfig}
\usepackage{dcolumn}
\usepackage{bm}
\newcommand{\be}{\begin{equation}}
\newcommand{\ee}{\end{equation}}
\newcommand{\bse}{\begin{subequations}}
\newcommand{\ese}{\end{subequations}}
\newcommand{\bary}{\begin{eqnarray}}
\newcommand{\eary}{\end{eqnarray}}
\newcommand{\bwt}{\begin{widetext}}
\newcommand{\ewt}{\end{widetext}}

\begin{document}


\title{Multi-TeV flaring from blazars: Markarian 421 as a case study}
\author{Sarira Sahu$^1$, Luis Salvador Miranda$^1$, Subhash Rajpoot$^2$ 
}

\affiliation{
$^1$Instituto de Ciencias Nucleares, Universidad Nacional Aut\'onoma de M\'exico, 
Circuito Exterior, C.U., A. Postal 70-543, 04510 Mexico DF, Mexico\\
$^2$Department of Physics \& Astronomy, California State
University,\\
1250 Bellflower Boulevard, Long Beach, CA, 90840, USA
}


\begin{abstract}

The TeV blazar Markarian 421 underwent multi-TeV flaring during
April 2004 and simultaneously observed in the
X-ray and TeV energies.  It was observed that the TeV outbursts had no
counterparts in the lower energy range, which implies that this might be
an orphan flare. We show that Fermi-accelerated protons of energy
$\leq 168$ TeV will interact with
the low energy tail of the background synchrotron self-Compton photons
in the inner region of the blazar to produce the multi-TeV flare and our
results fit very well the observed spectrum. Based on our study, we
predict that  the blazars with a deep valley in between the end of the
synchrotron spectrum and the beginning of the SSC spectrum are
possible candidates for orphan flaring. Future possible candidates for
this scanario are the HBLs Mrk 501 and PG 1553+113 objects.

\end{abstract}

\maketitle

\section{Introduction}

Active galactic nuclei (AGN) emit electromagnetic radiation 
from radio to gamma-rays and  exhibit large luminosity
variations on time scales ranging from less than an hour up to several
years.  A super massive black hole is believed to sit at the center of the AGN surrounded by 
an accretion disk in the inner region and a torus of gas cloud in the
outer region. Oppositely directed relativistic jets are ejected from
the AGN which are perpendicular to the accretion disk and the torus.
In the framework of the unification scheme of AGN,
blazars and radio galaxies are intrinsically the same objects,
viewed at different angles with respect to the jet axis. 
When the angle between the jet and the line of sight is small it is
called blazar and in contrast, for radio galaxies, the angle between
the jet and the light of sight is  large. Almost all AGN detected at
very high energy (VHE)  ($> 100$ GeV) are blazars with the exception
of the three objects, Centaurus A (Cen A)\cite{Abdo:2010fk,Roustazadeh:2011zz}, M87 and NGC 1275 which are
radio galaxies\cite{Ghisellini:1998it,Fossati:1998zn}. The spectral energy distribution (SED) of these AGN
have a double peak structure in the $\nu-\nu F_{\nu}$ plane. 
The low  energy peak corresponds to
the synchrotron radiation from a population of relativistic electrons
in the jet. Although the general consensus is that the high energy
peak corresponds to the synchrotron self
Compton (SSC) scattering of the high energy electrons with their
self-produced synchrotron photons, this result  remains inconclusive
for various reasons. However, the so called leptonic model is
very successful in explaining the multi wavelength emission from blazars and FR I
galaxies\cite{Fossati:1998zn,Ghisellini:1998it,Abdo:2010fk,Roustazadeh:2011zz}.

Although the SSC scenario seems to work very well to explain the SED
of AGN up to the second peak \cite{Dermer:1993cz,Sikora:1994zb}, difficulties arise in explaining the
multi-TeV emission detected in Cen A\cite{Aharonian:2009xn}, flares
from the radio galaxy M87 \cite{Abramowski:2011ze},
the  flares from blazars 1ES
1959+650\cite{Krawczynski:2003fq,Cui:2004wi} and  Markarian 421 (Mrk
421)\cite{Blazejowski:2005ih}. 
Also the inevitable outcome of
the leptonic scenario is that, emission in multi-TeV energy has to be
accompanied by a simultaneously enhanced emission in the synchrotron
peak. Unfortunately the enhanced synchrotron emission was not observed
in the flaring of 1ES 1959+650 in June 2002\cite{Krawczynski:2003fq} and  also probably
in the flaring of Mrk 421 in April 2004\cite{Blazejowski:2005ih},
which implies that the SSC model
may not be efficient enough to contribute in the multi-TeV regime. 

To explain the orphan flaring of 1ES1959+650 a hadronic synchrotron mirror model was
proposed by B\"ottcher\cite{Bottcher:2004qs}. In
this model, the high energy protons from the jet interact with the
primary  synchrotron photons that have
been reflected off clouds located at a few pc above the accretion
disk. These photons are blue shifted in the jet frame so that there
will be a substantial decrease in the high energy protons to overcome the
threshold for $\Delta$-resonance. Similarly a structured leptonic jet model is proposed
to explain the orphan  TeV flare\cite{Kusunose:2006gu}.

In the hadronic models, the second bump is believed to be formed due
to the synchrotron emission from the ultra-high energy protons, or
emission from the secondary electron-positron pairs or muons produced
from the charged pion decay as a reasult of the interaction of the high energy
protons with the background low energy photons.  In this process the
decay of neutral pions to $\gamma$-ray pairs can explain the high
energy peak in the multi-TeV range.

Previously  it was shown that the above hadronic processes are inefficient and
to explain the high energy peaks efficient accleration of relativistic protons to ultra-high
energies within the inner part of the jet outflow is required. At the same time
the jet kinetic power has to exceed the Eddington luminosity by orders
of magnitude. The hadronic model is employed by Zdziarski et al.\cite{Zdziarski:2015rsa} to
explain the broad band spectra of radio-loud active galactic
nuclei and by Cao et al.\cite{Cao:2014nia}  to explain the TeV spectrum of the
blazar 1ES 1101-232. In both these cases it is shown that super
Eddington luminosity in proton is required to explain the high enery
peaks.  This situation arises because the photon density in the jet is
low.  However, this problem can be circumvented if one assumes that
within the jet there is a dense inner jet region\cite{Sahu:2013ixa}.
The multi-TeV peak  can also be produced
by $pp$ interactions. It was shown earlier that normally the $pp$ process is inefficient to
produce $\gamma$-rays in the blazar environment unless the 
jet-cloud interaction is taken into 
account\cite{Atoyan:2001ey,Atoyan:2002gu,Dermer:2012rg}.
The emisson of gamma-rays by the interaction of a population of
massive stars surrounding the AGN jets is studied in\cite{Bednarek:1996ffa,Araudo:2013ora}. Also emission of
high energy gamma-rays by the interaction of
the AGN jets with the tidally disrupted atmosphere of  Red Giant is
studied in\cite{ Barkov:2010ew,BoschRamon:2012td}.

\section{Photohadronic Model}

To address the orphan TeV flaring of 1ES 1959+650 and the
multi-TeV emission from Cen A and M87, Sahu et
al. \cite{Sahu:2013ixa,Sahu:2012wv,Sahu:2013cja} have used the
hadronic model. Recently this model is also used to estimate the
neutrinos flux from the TeV-blazars and their spatial correlation
with the IceCube events\cite{Sahu:2014fua}. Here we use the same
hadronic model to explain the multi-TeV orphan flaring of Mrk 421. In
this model the 
Fermi-accelerated high energy protons interact 
with the SSC photons in the core region of the jet (few times the
Schwarzschild radius $R_S$)  to produce the
$\Delta$-resonance. Subsequently the $\Delta$-resonance decays to
charged and neutral pions as follows,
\be
p+\gamma \rightarrow \Delta^+\rightarrow  
 \left\{ 
\begin{array}{l l}
 p\,\pi^0, & \quad 
\\
  n\,\pi^+  \rightarrow n e^{+}\nu_e\nu_\mu \bar\nu_\mu
& \quad  
\\
\end{array} \right. .
\label{decaymode}
\ee
The decay of neutral pions to TeV photons gives the multi-TeV SED. 
Throughout our work we use natural units $c=\hbar=1$.
The relationship between the $\pi^0$-decay TeV photon energy
$E_{\gamma}$ and the 
target SSC photon energy $\epsilon_{\gamma}$ in the observer frame is given by
\be
E_\gamma \epsilon_\gamma \simeq 0.032~\frac{ {\cal D}^2}{(1+z)^2} ~{\rm GeV}^2,
\label{Eegamma} 
\ee
where ${\cal D}$ is the Doppler factor and $z$ is the redshift. The
observed TeV $\gamma$-ray energy and the proton energy
$E_p$ are related through
\be
E_p=\frac{10\Gamma}{\cal D} E_{\gamma}, 
\label{Eproton}
\ee
where $\Gamma$ is the bulk Lorentz factor of the relativistic
jet. 

\begin{figure}[t!]
\vspace*{-0.3cm}
{\centering
\resizebox*{0.5\textwidth}{0.4\textheight}
{\includegraphics{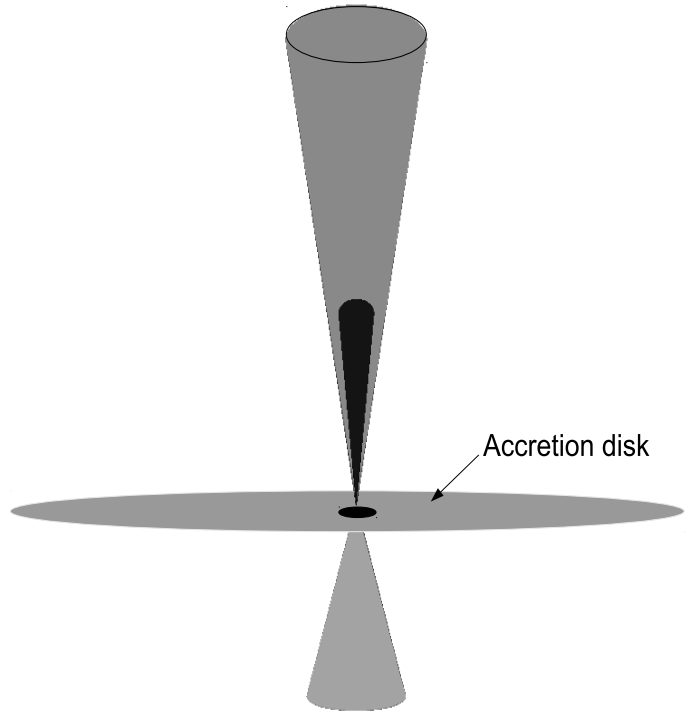}}
\par}
\vspace*{0.30cm}
\caption{Geometry of the orphan flaring of blazar Mrk 421: the hidden interior
  compact cone (jet) is responsible for the multi-TeV orphan flaring and the exterior cone
  corresponds to the normal jet.} 
\label{geometrym421}
\end{figure}

Multi-TeV emission is observed from many blazars and FR I galaxies by
Cherenkov telescope arrays. These emission  
are classified into non-flaring and flaring events.
For the non-flaring events (e.g. emission from Cen
A\cite{Abdo:2010fk}), the injected proton spectrum is a power-law 
given by $dN_p/dE_p\propto E_p^{-\alpha}$ with $\alpha \ge 2$. These protons will interact
with the background SSC photons having comoving number density
$n'_{\gamma}$ (henceforth $^{\prime}$ implies the jet comoving frame) and satisfying the kinematical
conditions in Eqs.(\ref{Eegamma})  and (\ref{Eproton}).
As discussed in Ref.\cite{Sahu:2013ixa}, the flaring occurs within a
compact and confined volume of a smaller cone which is enclosed in a
bigger cone as shown in Fig. \ref{geometrym421}. 
The internal compact jet models are also proposed to explain the fast
variability of the blazars Mrk 501 and PKS 2155-304\cite{Giannios:2009kh,Ghisellini:2008us}.
In our case, 
the injected proton spectrum is a power-law with an exponential decay.
In a unified manner we can express the injected proton spectrum for both non-flaring and
flaring  processes as\cite{Sahu:2013ixa,Aharonian:2003be}
\bary
\frac{dN_p}{dE_p}\propto E_p^{-\alpha}\left\{
\begin{array}{l l}
1, & \quad\text{non-flaring}\\
e^{-E_p/E_{p,c}}, & \quad \text{flaring}
\end{array} \right. ,
\eary
where $E_{p,c}$ is the break energy for the high energy protons and
above this break energy the proton spectrum falls very fast. The high energy
protons will interact in the flaring region having the comoving photon
number density $n'_{\gamma,f}$ to produce the $\Delta$-resonance.
The photon density in the flaring region is much higher than
the rest of the blob (non-flaring region with the photon density $n'_{\gamma}$) probably due to the copious
 annihilation of electron positron pairs, splitting of photons in the
 magnetic field, enhanced IC photons in this region and Poynting flux dominated flow
which can form from the magnetic  reconnection in the strongly
magnetized plasma around the base of the jet\cite{Kachelriess:2010zk,Giannios:2009kh,Giannios:2009pi}.
Here we assume that the spectrum has a exponential decay above the $E_{p,c}$ only for the
 flaring case. In this model
 the inner compact region has a size $R'_f$ smaller than
 the size of the outer region  $R'_b$ i.e. $R'_f < R'_b$. 
Unfortunately, we do not know exactly the properties of the inner jet 
except  for the number density, the energy density of photons and and
the magnetic field are higher than  the corresponding values in the
outer jet. Due to higher photon density in the inner region, the multi-TeV
$\gamma$-rays and neutrinos can be produced through intermediate
$\Delta$-resonance as shown in Eq. (\ref{decaymode}). The
optical depth of the $\Delta$-resonance process in the inner jet region
is given by
\be
\tau_{p\gamma}=n'_{\gamma, f} \sigma_{\Delta} R'_f.
\label{optdepth}
\ee
The efficiency of the  $p\gamma$ process depends on the
physical conditions of the interaction region, such as the size, 
distance from the base of the jet, photon density and their
distribution in the region.

In the inner region we  compare the dynamical time scale $t'_{d}=R'_f$ 
with the $p\gamma$ interaction time scale
$t'_{p\gamma}=(n'_{\gamma,f}\sigma_{\Delta} K_{p\gamma})^{-1}$ to
constraint the seed photon density so that multi-TeV photons can be
produced. For a moderate efficiency of this process, we can assume
$t'_{p\gamma} > t'_{d}$ and this gives
$\tau_{p\gamma} < 2$, where we take the inelasticity parameter
$K_{p\gamma}=0.5$. 
Also by assuming the Eddington luminosity is equally shared by the jet
and the counter jet, the luminosity within the inner region for a seed
photon energy $\epsilon'_{\gamma}$ will satisfy $(4\pi n'_{\gamma,f}
R'_f \epsilon'_{\gamma}) \ll L_{Edd}/2$.  This puts an upper limit on
the seed photon density as 
\be
n'_{\gamma,f}\ll \frac{L_{Edd}} {8\pi R'^2_f
\epsilon'_{\gamma}}.
\label{nedd}
\ee
From Eq.(\ref{nedd})  we can estimate the photon density in this
region and this is discussed in Sec. 3.
On the other hand,
the photon density in the outer region can be calculated from the
observed flux. For this we assume a scaling behavior of
 the photon densities in the flaring and the non-flaring regions which
 we take to be the following,,
\be
\frac{n'_{\gamma, f}(\epsilon'_{\gamma_1})}
{n'_{\gamma, f}(\epsilon'_{\gamma_2})}=\frac{n'_\gamma(\epsilon'_{\gamma_1})}
{n'_\gamma(\epsilon'_{\gamma_2})},
\label{denscale}
\ee 
this implies that, the ratio of photon densities at two different
background energies $\epsilon'_{\gamma_1} $  and $\epsilon'_{\gamma_2} $ 
 in flaring and non-flaring states remains the same. 
For a self consistent treatment, in principle we should use the photon density $n'_{\gamma,f}$ in the
hidden internal jet and solve the coupled transport equations
for leptons and photons along the jet axis by taking into account their
respective cooling mechanisms as well as the injection spectrum of the
primary particles\cite{Reynoso:2012wx}.
To avoid this complication we assume the scaling behavior of the photon
densities in different background energies as shown in
Eq. (\ref{denscale}). 
In the photohadronic scenario, the number of $\pi^0$-decay photons at a given energy are proportional
to both the number of high energy protons and the density of the SSC background
photons in the jet, i.e. $N(E_{\gamma})\propto N(E_p)
n'_{\gamma}$. For the flaring
case $n'_{\gamma}$ is replaced by the photon density in the flaring
region given by $n'_{\gamma,f}$. 
The $\gamma$-ray flux from the $\pi^0$ decay is then given by
\be
F_{\gamma}(E_{\gamma}) \equiv E^2_{\gamma} \frac{dN(E_\gamma)}{dE_\gamma} 
\propto  E^2_p \frac{dN(E_p)}{dE_p} n'_{\gamma,f} .
\ee
Using the scaling behavior of Eq.( \ref{denscale}),  the observed multi-TeV photon flux from $\pi^0$-decay
at two different observed photon energies $E_{\gamma 1}$ and
$E_{\gamma 2}$ can be expressed as
\be
\frac{F_\gamma(E_{\gamma_1})}{F_\gamma(E_{\gamma_2})} 
= 
\frac{n'_\gamma(\epsilon_{\gamma_1})}
{n'_\gamma(\epsilon_{\gamma_2})}
\left(\frac{E_{\gamma_1}}{E_{\gamma_2}}\right)^{-\alpha+2}
e^{-(E_{\gamma_1}-E_{\gamma_2})/E_c},
\label{denspectrum}
\ee
where $E_{\gamma_{1,2}}$ correspond to the proton energy
$E_{p_{1,2}}$. In this derivation we
have used the relations $E_{p_1}/E_{p_2}=E_{\gamma_1}/E_{\gamma_2}$, and
$E_{p_{1,2}}/E_{p,c}=E_{\gamma_{1,2}}/E_c$. By taking the ratio we get
rid of the proportionality constant.
By using the known flux at a particular energy
in the flaring/non-flaring state, we can calculate the  flux at other
energies using Eq.(\ref{denspectrum}) and also the normalization
constant can be calculated.

In terms of SSC photon energy and its luminosity, the photon number density
$n^{\prime}_{\gamma}$ is expressed as
\be
n^{\prime}_{\gamma}(\epsilon_{\gamma}) = \eta \frac{L_{\gamma,
    SSC} (1+z)}{{\cal D}^{2+\kappa} 4\pi {R^{\prime}}^2_b \epsilon_{\gamma}},
\ee
where $\eta\sim 1$ is the efficiency of SSC process and $\kappa\sim
(0-1)$ depending on whether the jet is continuous or discrete. In this
work we consider  $\kappa=0$. 
The SSC photon luminosity  is expressed in terms of observed flux
 ($\Phi_{SSC}(\epsilon_{\gamma}) =\epsilon^2_{\gamma} dN_{\gamma}/d\epsilon_{\gamma}$)  which is
given by
\be
L_{\gamma,SSC}=\frac{4\pi d^2_L \Phi_{SSC}(\epsilon_{\gamma})}{(1+z)^2}.
\ee
Furthermore, by using Eq.(\ref{Eegamma}), we can simplify the
ratio of photon densities given in Eq.(\ref{denscale}) to
\be
\frac{n'_\gamma(\epsilon_{\gamma_1})}
{n'_\gamma(\epsilon_{\gamma_2})}=\frac{\Phi_{SSC}(\epsilon_{\gamma
    1})}{\Phi_{SSC}(\epsilon_{\gamma 2})}
 \frac{E_{\gamma_1}}{E_{\gamma_2}}.
\ee
Now we can express Eq.(\ref{denspectrum}) in terms of observed SSC
flux and $E_{\gamma}$ as,
\be
\frac{F_\gamma(E_{\gamma_1})}{F_\gamma(E_{\gamma_2})} 
=
\frac{\Phi_{SSC}(\epsilon_{\gamma1})}{\Phi_{SSC}(\epsilon_{\gamma 2})}
\left(\frac{E_{\gamma_1}}{E_{\gamma_2}}\right)^{-\alpha+3}
e^{-(E_{\gamma_1}-E_{\gamma_2})/E_c}.
\label{sscspectrum}
\ee
Although Eqs.({\ref{denspectrum}) and ({\ref{sscspectrum}) give the
same result, the latter form is much more simpler than the former one.
The one in Eq.({\ref{sscspectrum}) uses the SED of SSC photon calculated
using the leptonic model. Here the multi-TeV flux is proportional to
$E_{\gamma}^{-\alpha+3}$  and $\Phi_{SSC}(\epsilon_{\gamma})$ while in the former case it is proportional
to $E_{\gamma}^{-\alpha+2}$ and to the photon number density $n^{\prime}_{\gamma}(\epsilon_{\gamma})$.
Finally we would like to point out that in the photohadronic process ($p\gamma$), the
multi-TeV photon flux can be expressed as
\be
F(E_{\gamma})=A_{\gamma} \Phi_{SSC}(\epsilon_{\gamma} )\left (
  \frac{E_{\gamma}}{TeV}  \right )^{-\alpha+3}e^{-E_{\gamma}/E_c},
\label{modifiedsed}
\ee
where $A_{\gamma}$ is a dimensionless constant and both
$\epsilon_{\gamma}$ and $E_{\gamma}$ satisfy the condition given in
Eq.(\ref{Eegamma}). This formula will be used to calculate the
multi-TeV flux from both non-flaring (without exponential decay term)  and flaring events from AGN and
their subclass if the emission is due to photohadronic process from
the core region. For different blazars/AGN, the value of $A_{\gamma}$
will be different which we shall discuss in the next section.
We can calculate the Fermi accelerated high energy proton flux $F_p$
from the TeV $\gamma$-ray flux through the relation
\be
F_p(E_p)=7.5\times \frac{F_{\gamma}(E_{\gamma})}{\tau_{p\gamma}(E_p)}.
\label{pflux}
\ee
The optical depth $\tau_{p\gamma}$ is given in
Eq.(\ref{optdepth}). For the observed highest energy proton energy
$E_p$, $F_p(E_p)$ will be smaller than the Eddington flux
$F_{Edd}$. This condition puts a lower limit on the optical depth of
the process and is given by
\be
\tau_{p\gamma} (E_p) > 7.5\times
\frac{F_{\gamma}(E_{\gamma})}{F_{Edd}}.
\label{optdepth}
\ee
From the comparison of different times scales and from
Eq.(\ref{optdepth}) we will be able to constraint the seed photon
density in the inner jet region.

\begin{figure}[t!]
\vspace*{-0.2cm}
{\centering
\resizebox*{0.5\textwidth}{0.4\textheight}
{\includegraphics{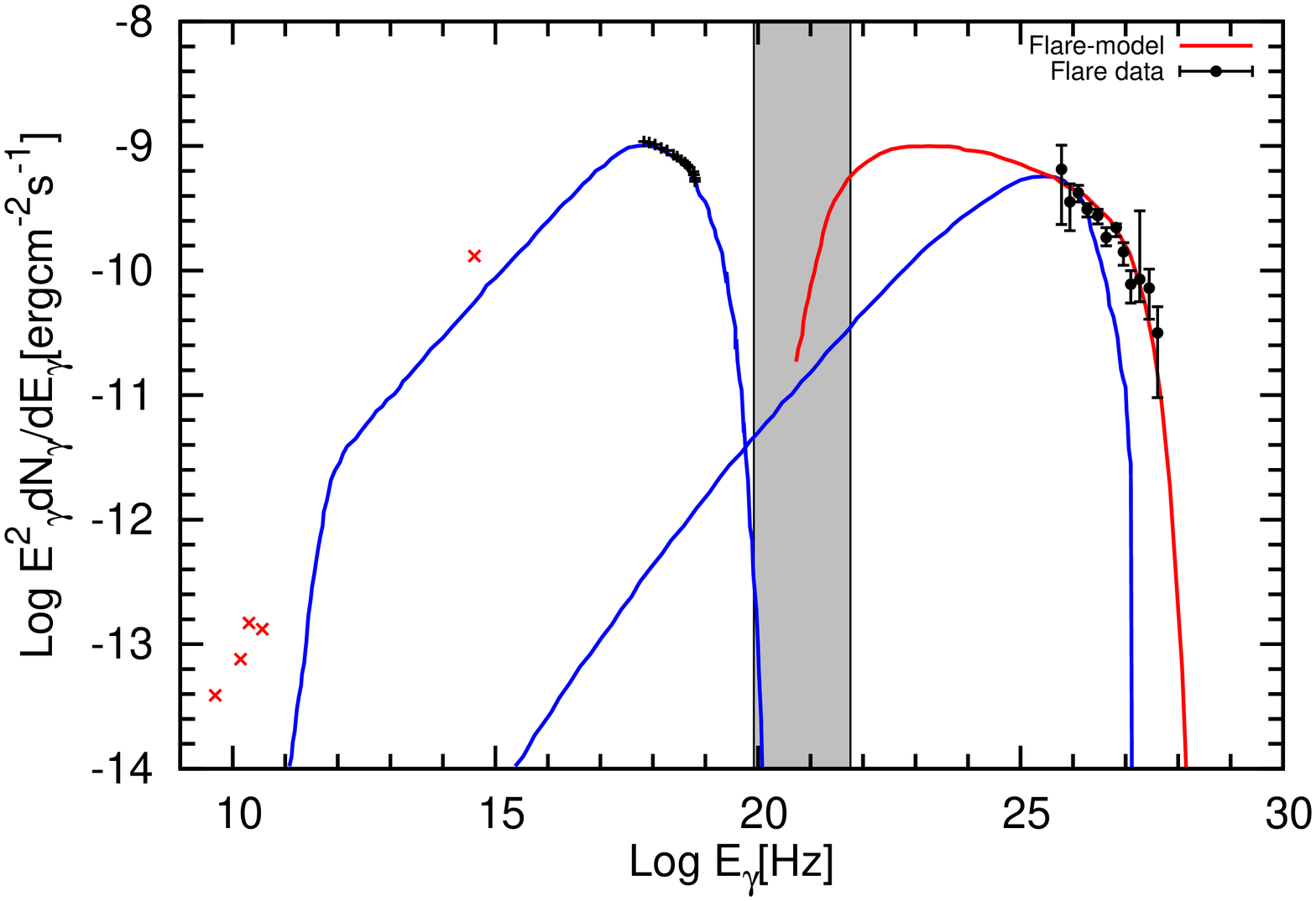}}
\par}
\vspace*{-0.30cm}
\caption{The SED of Mrk 421 is shown in all the energy bands which are
  taken from Ref.\cite{Blazejowski:2005ih}.
The flare of April 2004 in multi-TeV energy is shown here. The
hadronic model fit to the April 2004 data is shown as continuous line to the extreme
right. The shaded region is the energy range of SSC photons where the
Fermi-accelerated protons are collided to produce the $\Delta$-resonance.
}
\label{m421sed}
\end{figure}

\begin{figure}[t!]
\vspace*{-0.2cm}
{\centering
\resizebox*{0.5\textwidth}{0.4\textheight}
{\includegraphics{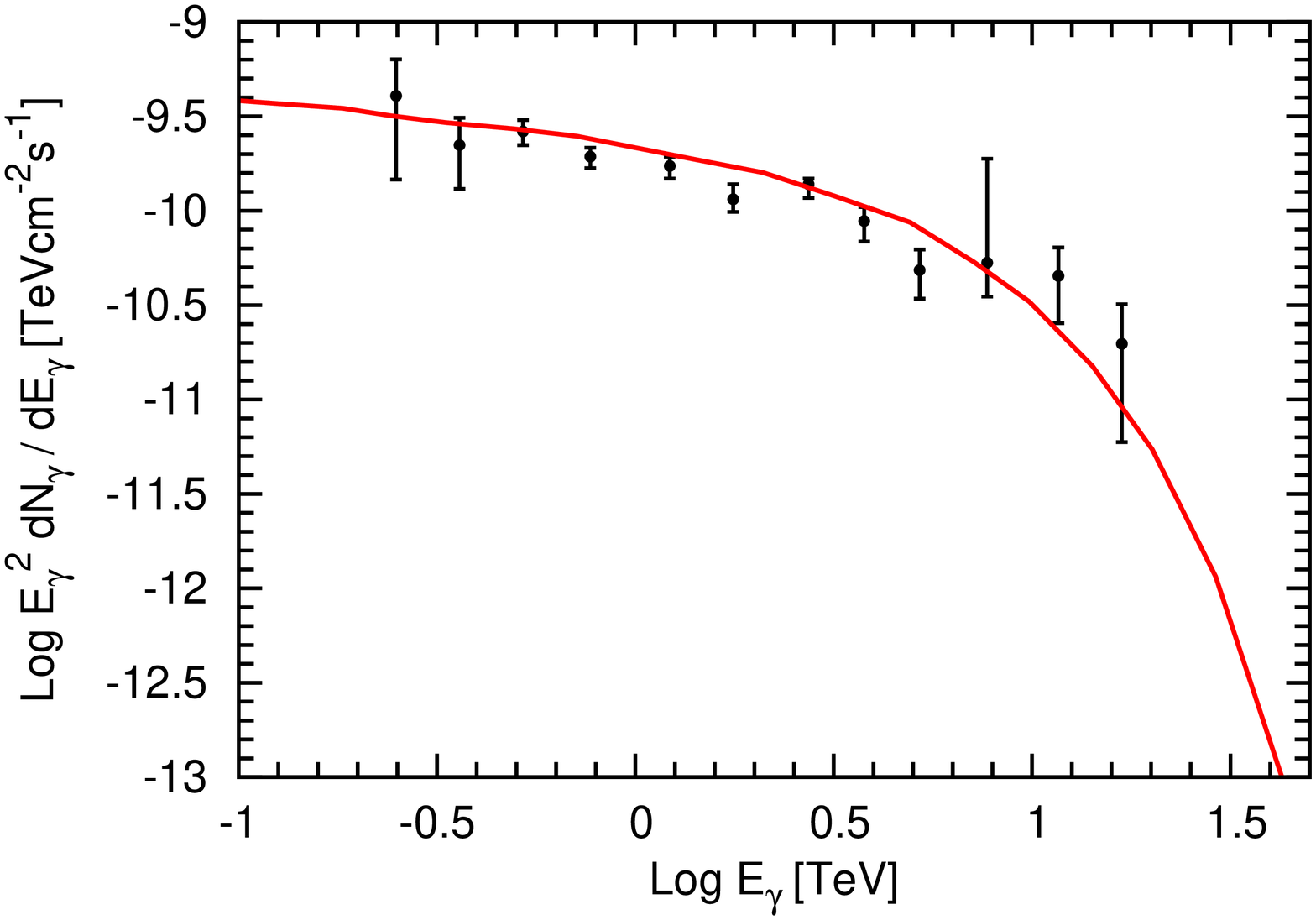}}
\par}
\vspace*{-0.30cm}
\caption{The continuous curve is the hadronic model fit to the
  multi-TeV flaring data of Mrk 421.
}
\label{m421tevsed}
\end{figure}


\section{Mrk 421}

Mrk 421 is a high synchrotron peaked BL Lac object
(HBL) and is the first extragalactic source with a redshift of z=0.031
to be established as a TeV emitter\cite{Punch:1992xw}. It has a luminosity distance $d_L$
of about 129.8 Mpc. Its central supermassive black
hole is asumed to have  a mass $M_{BH}\simeq (2-9)\times 10^8\, M_{\odot}$
corresponding to a Schwarzschild radius of $(0.6-2.7)\times 10^{14}$
cm and the Eddington luminosity $L_{Edd}=(2.5-11.3)\times 10^{46}\, erg\,s^{-1}$. 
The synchrotron peak of its SED is in the soft to medium X-ray range and the
SSC peak is in the GeV range. It is one of the fastest varying
$\gamma$-ray sources. In the past through dedicated multi wavelength
observations, the source has been studied intensively. These studies
show a correlation between X-rays and very high energy (VHE)
$\gamma$-rays. A one-zone SSC model explains the observed SED
reasonably well\cite{Abdo:2011zz}. Several major flares were observed in
the 2003/2004 season. 
During April 2004, a large flare took place both in the X-rays and the
TeV
energy band. The flare lasted for more than two weeks (from MJD 53,104
to roughly MJD 53,120). But due to a large data gap between MJD 53,093
and 53,104, it is difficult to exactly quantify the duration. The source was observed
simultaneously at TeV energies with the Whipple 10 m telescope and at
X-ray energies with the Rossi X-ray Timing Explorer (RXTE). It was
also observed simultaneously in radio and optical wavelengths. 
During the flaring it was observed that, the TeV flares had no
coincident counterparts at longer wavelengths. Also
it was observed that the X-ray flux reached its peak 1.5 days before the
TeV flux did during this outburst. So it is believed that, the TeV flare
might not be a true orphan flare like the one observed in 1ES
1959+650.  On the other hand  remarkable similarities between
the orphan TeV flare in 1ES 1959+650 and Mrk 421 were observed,
including similar variation patterns in X-rays.

\begin{table}
\centering
\caption{These parameters (up to $B'$) are taken from the one-zone synchrotron
model of ref. \cite{Blazejowski:2005ih} which are used to fit the SED of
Mrk 421. The last three parameters are obtained from the best
fit to the observed flare data  in our model.} 
\label{tab1}
\begin{tabular*}{\columnwidth}{@{\extracolsep{\fill}}llll@{}}
\hline
\multicolumn{1}{@{}l}{Parameter} &Description & Value\\
\hline
$M_{BH}$ & Black hole mass & $(2-9)\times 10^8 M_{\odot}$\\
z & Redshift & 0.031\\
$\Gamma$ &Bulk Lorentz Factor & 14\\
${\cal D}$& Doppler Factor & 14\\
$R^{\prime}_b$ & Blob Radius & $0.7\times 10^{16}$cm\\ 
$B^{\prime}$ &Magnetic Field & $0.26$ G\\ 
\hline
$R'_f$ &Inner blob Radius& $3\times 10^{15}$cm\\
$\alpha$ &Spectral index& $2.7$\\
$E_c$ &$\gamma$-ray Cut-off Energy& $6.2$ TeV\\
\hline
\end{tabular*}
\end{table}

By using the one-zone SSC model, the average SED of Mrk 421 is
fitted in Fig. 11 of ref. \cite{Blazejowski:2005ih}. 
In Fig. \ref{m421sed} we have shown the SED of Mrk 421. In this
figure, the red crosses to the extreme left are the measurements 
in the radio frequency
band measured by 26 m telescope at the University of
Michigan Radio Astronomy Observatory (UMRAO) and by the 13.7 m
Mets\"ahovi radio telescope at Helsinki. The single cross (red) in the optical range is measured by
Fred Lawrence Observatory (FLWO) 1.2 m telescope. The points of the
first peak in the X-ray range are from Rossi X-ray Explorer (RXTE)\cite{Blazejowski:2005ih}.

In the one-zone leptonic model,
the blob of size $R^{\prime}_b\sim 0.7\times 10^{16}\, cm$ moves down
the conical jet with a Lorentz factor $\Gamma\simeq 14$ and a Doppler
factor of ${\cal D}=14$. The emitting region is filled with an
isotropic electron population and a randomly oriented magnetic field
$B^{\prime}=0.26$ G. In the present work we study the flaring of Mrk
421 during April 2004. We use the parameters of the one-zone leptonic model of
ref. \cite{Blazejowski:2005ih}. The parameters of the one-zone
synchrotron model are summarized in Table \ref{tab1}. In principle the
Lorentz factor in the inner jet should be larger than the outer
jet. But here we assume that $\Gamma_{out}\simeq \Gamma_{in}\simeq \Gamma$.

\section{Results}

  The flaring of Mrk 421 in April 2004 was observed in the energy
range $ 0.25\, TeV (6.0\times 10^{25}\, Hz) \le E_{\gamma} \le
16.85\, TeV  (4.1\times 10^{27}\, Hz) $ by the Whipple 
telescope. In the hadronic model discussed above, this corresponds
to the Fermi accelerated proton energy  in the range $2.5\, TeV \le
E_p \le 168\, TeV$ and the corresponding background photon energy will
lie in the range $23.6\, MeV (5.7\times 10^{21}\, Hz) \ge
\epsilon_{\gamma} \ge 0.35\, MeV (8.4\times 10^{19}\, Hz) $. 
This range of $\epsilon_{\gamma}$ is the  shaded region shown in
Fig. \ref{m421sed} which is in the low energy tail of the SSC
photons. 
For the calculation of normalized
multi-TeV flux we take into account one of the observed TeV fluxes
from the flare with its corresponding energy and with the use of 
Eq.({\ref{sscspectrum}) calculate other TeV fluxes. 
In this model the free parameters are the spectral index
$\alpha$ and the TeV $\gamma$-ray cut-off energy $E_c$ which are adjusted to obtain
the best fit and the values are $\alpha=2.7$ and
$E_c=6.2$ TeV. This value of $E_c$ corresponds to the proton cut-off energy
$E_{p,c}=62$ TeV and the background SSC photon enery $\epsilon_{\gamma,SSC}=0.96\,
MeV$ ($2.3\times 10^{20}\, Hz$) which is very close to the beginning of the SSC
energy as shown in Fig. \ref{m421sed}. 
This shows that for orphan flaring, the cut-off energy
$E_c$ is due to the change from the synchrotron band to the  SSC
band and can be calculated from their crossover energy.

In Fig. \ref{m421tevsed}  we show both the observed
multi-TeV SED and our model fit to it. In our results, the presence of
  $\Phi_{SSC}(\epsilon_{\gamma})$ in Eq. (\ref{modifiedsed}) modifies
  the power-law with the exponential fall scenario. From the best fit
  parameters we obtain the value of the dimensionless constant
  $A_{\gamma}\simeq 20$ in Eq. (\ref{modifiedsed}). By using the best
  fit parameters, we have also calculated
  the value of $A_{\gamma}$  from the multi-TeV flare of 1ES 1959+650
  and M87 and the multi-TeV emission from Centaurus A, which are given
  as 86, 1.86 and $8.6\times 10^{-4}$ respectively. We observe that
  for orphan flaring the condition $A_{\gamma}\gg 1$ is satisfied as
  is seen from 1ES 1959+650 and Mrk421. On the other hand for
  non-orphan flaring this value is small i.e. $A_{\gamma} \leq 1$.

In the flaring state, the proton luminosity $L_p$ for the highest
observed proton energy $E_p=168$ TeV has to be smaller than the
$L_{Edd}\sim 2.5\times 10^{46}\, erg\, s^{-1}$ and this gives $\tau_{p\gamma} > 0.02$ in the inner
jet region. For our estimates
we consider the hidden jet size $R'_f\simeq 3\times 10^{15}$ cm which
is between $R_s$ and the blob radius $R'_b$. This value of $R'_f$
corresponds to a day scale variability. We take day scale variability
due to the fact that the flaring lasted for more than two weeks.
If we consider small $R'_f$
then the seed photon density will increase as can be seen from
(\ref{nedd}). But this increase in density
will not affect our
results because we are using the scaling behavior of Eq.(\ref{denscale}). The constraint
on the optical depth gives the lower limit on the seed photon density
in the inner region $n'_{\gamma, f} > 1.3\times 10^{10}\,
cm^{-3}$. Again by assuming $t'_{p\gamma} < t'_d$ we obtain the upper
limit on the optical depth $\tau_{p\gamma} < 2$ and this corresponds
to the photon density $n'_{\gamma,f} < 1.3\times 10^{12} cm^{-3}$. We
have also estimated the seed photon density from Eq.(\ref{nedd}), for
$\epsilon_{\gamma}=0.35$ MeV, which gives $n'_{\gamma,f} <8.9\times
10^{10} cm^{-3}$. The upshot of this analysis is that we get the
constraint $1.3\times 10^{10}\,
cm^{-3} < n'_{\gamma,f} < 8.9\times 10^{10} cm^{-3}$, which shows that the photon
density in this region is high. This range of photon density
corresponds to the optical depth in the range $0.02 < \tau_{p\gamma} <
0.13$ and the proton flux at $E_p=168$ TeV is below the
$F_{Edd}\sim 1.24\times 10^{-8}\, erg\,cm^{-2}\,s^{-1}$. 
Due to the adabatic expansion of the inner blob,
the photon density will be reduced to $n'_{\gamma}$ and the energy
will dissipate once these photons are in the
bigger cone. So even if we have two-zones (the inner and the outer),
only the outer zone will be responsible for the observed synchrotron
and IC peaks.  From the leptonic model fit to the SED, the magnetic
field in the outer jet region is $B'=0.26$ G and higher magnetic field
is expected in the inner jet region. The maximum proton energy in the
hidden jet region will be 
$E_{p,max}\sim 10^{18} (B'_f/1 G)$ eV and for larger magnetic
field the $E_{p,max}$ can even be higher.

In the normal jet scanario we estimate the proton flux needed to explain
the observed TeV $\gamma$-rays. Corresponding to
the background energy $\epsilon_{\gamma}=0.35$ the
photon density is $n'_{\gamma}\sim 4 \times 10^{3}\, cm^{-3}$ and the
the optical depth is $\tau_{p\gamma}\sim 1.4\times 10^{-8}$. 
As discussed above, for background SSC photon energy 0.35 MeV, 
the observed TeV photon has the energy 
$E_{\gamma}=16.85$ TeV and its flux is $F_{\gamma}  \sim 3.16\times 10^{-11}\, erg\, s^{-}\,
cm^{-2}$. The high energy proton responsible for the photohadronic
process to produve this TeV $\gamma$-ray has energy $E_p=168$ TeV and we can
estimate its flux from Eq.(\ref{pflux}) which gives $F_p\simeq
10^6\times F_{Edd}$. This shows that the normal jet model needs super
Eddington power in protons to explain the high energy peaks, whereas,
the inner jet scenario exterminates this extreme energy requirement.

The high energy protons will be accompanied by high energy electrons
in the same energy range. In the magnetic field of the jet, these
electrons will emit synchrotron photons in the energy range $4\times
10^{19}\, Hz$ to $2\times 10^{23}\, Hz$, which is in the lower part of
the SSC spectrum and will not be observable because of the lower flux.
Also the SSC emission will take place from these high energy electrons
and the energy of these IC photons will be $E_{IC}\sim \gamma^2_e
\epsilon_{syn}$. By considering the electron Lorentz factor in the range
$7\times 10^2\le \gamma_e\le 4\times 10^4$\cite{Abdo:2011zz}
and the peak energy of the synchrotron photon $\epsilon_{syn}\sim 10^{18}\,
Hz$, the SSC process can contribute in the energy range $2\, GeV \le
E_{IC}\le 6.6\, TeV$. But the details of the SSC flux
depends on the breaks in SSC spectrum and the spectral index. 
It is observed that during the flaring of Mrk 421, the X-ray emission
reached the peak, days after the TeV emission, which poses a serious
challange to the SSC model\cite{Blazejowski:2005ih}. 
The SSC contribution to the
multi-TeV band will be very much supressed.
Similarly the multi-TeV photons in the energy range $0.25\, TeV
\leq E_{\gamma} \leq 16.85\, TeV$ can interact with the
background photons to produce $e^+e^-$ pair and these individual
electron or positron will have energy $E_{\gamma}/2$. To produce
$e^+e^-$ pair the required threshold seed photon energy 
$\epsilon_{\gamma}\geq 2 m^2_e/E_{\gamma}$ is needed.
During the flaring, the multi-TeV $\gamma$-rays in the energy range
$0.25\, TeV\leq E_{\gamma} \leq 16.85\, TeV$ will 
interact with the soft seed photons in the energy range $0.05\, eV\leq
\epsilon_{\gamma}\leq 3.5\, eV$ (in between the infrared and the visible range), where
$\sigma_{\gamma\gamma}\sim 1.7\times 10^{-25}\, cm^2$ is the maximum
cross section and for higher
$\epsilon_{\gamma}$,  the $\sigma_{\gamma\gamma}$ will be smaller. The origin
of these soft photons are from the synchrotron emission of $1-10$ GeV
electrons in a magnetic field $\sim 1$ Gauss towards the base of the
evolving jet as well as the ambient photons from the disk. 
On the other hand multi-TeV $\gamma$-rays are produced
beyond this region where the photons are in the low energy tail
($0.35\, MeV - 23.5\, MeV$ range) of the IC photons. 
These two regions are distinct and there will not be enough
low energy seed photons (0.05-3.5 eV) in the IC tail region. 
So, mostly the TeV photons will encounter the IC photons and the pair
production cross section for $\epsilon_{\gamma} \geq 0.35\, MeV$ is
very small $\sigma_{\gamma\gamma} \leq 10^{-30}\, cm^2$, which
corresponds to amean  free path $\lambda_{\gamma\gamma} \geq 10^{19}\,cm$.
Hence, TeV photons will not be attenuated much due to $e^+e^-$ pair
production.  Also it has been observed that, during the flaring of Mrk 421, the
variation in the light curves at optical and radio wavelengths are
slight\cite{Blazejowski:2005ih},
which shows that the low energy photon production was
suppressed. The positron produced from the $\pi^+$ decay will have energy
$E_{\gamma}/2$ and it will radiate synchrotron photons in the
energy range $2\times 10^{17}\, Hz$ to $9\times 10^{20}\, Hz$. The
photon flux $F_{e^+,syn}$ from the  synchrotron radiation of $e^+$ 
will be much smaller than $F_{\gamma}(E_{\gamma}=0.25\,
TeV)/8$, i.e., $F_{e^+, syn} \ll 8\times 10^{-11}\, erg\,
cm^{-2}\, s^{-1}$. This flux is well below the observed flux limit in the
normal case as can be seen from Fig. \ref{m421sed}.
From the above analysis our conclusion is that, the photon fluxes from
the synchrotron emission of the electrons and positrons are not
observable during the flaring event of  Mrk 421 in April 2004, which
makes this flare orphan, like the one observed in 1ES 1959+650.
In principle, the multi-TeV $\gamma$-rays from the extragalactic
sources can be reduced due to
the absorption of TeV photons by the diffuse extragalactic background
light (EBL) through the process
$\gamma_{\rm TeV}+\gamma_{\rm b}\rightarrow e^+e^-$ due to the energy
dependent optical depth\cite{Finke:2009xi,Dominguez:2010bv,Inoue:2012bk}.
But for low redshifts and the energy rage of our interest, the optical depth does not vary
much. Hence we can assume almost a constant optical
depth\cite{Aharonian:2003be} so that the spectral
shape remains nearly unchanged.

In conclusion, from the study of the flaring events in blazars, we deduce that
the flaring phenomena can be explained through the
photohadronic interaction in a compact and confined region within the
blazar jet where the photon density is high. From the study of 1ES 1959+650 and Mrk 421, we deduce that
orphan flaring can only be possible for those blazars
which have a deep valley in between the end of the synchrotron SED and
the beginning of the SSC SED as shown in Fig. \ref{m421sed}.
 We note that the HBLs Mrk 501 and PG 1553+113 are possible candidates
for orphan flaring in future. For the orphan flaring events we find
$A_{\gamma} \gg 1$ and for non-orphan flaring $A_{\gamma} \leq 1$.

If Mrk 501 were to produce an orphan flare then the flare energy will
lie in the range $1\, TeV \leq E_{\gamma}\leq 8.6\, TeV$ and
this corresponds to the background SSC photon energy in the range 
$4.3\, MeV \geq \epsilon_{\gamma} \ge 0.5\, MeV$. We estimate
this by taking the parameters of Mrk 501 as follows: ${\cal D}=12$,
$z=0.034$ from Ref.\cite{Collaboration:2012wv}. In this case the
maximum Fermi-accelerated proton energy will be 
$E_p \le 10E_{\gamma}\sim 86\, TeV$. Also, another condition which
must hold for the orphan flaring is 
\be
A_{\gamma}=\frac{F(8.6\, TeV)}{\Phi_{SSC}(0.5\, MeV)} \left (
  \frac{1}{8.6} \right )^{-\alpha+3} e^{8.6/(E_c/TeV)} \gg 1.
\label{condflare}
\ee
It has to noted that, if ${\cal D}$ changes, accordingly
the values of $E_{\gamma}$, $\epsilon_{\gamma}$ and the $E_p$ will
also change, but the condition of Eq.(\ref{condflare}) is independent
of these changes.

\section{Conclusions}

The orphan flaring of Mrk 421 can be explained well by the hadronic
model. We observe that, in this model, the multi-TeV photon flux is
proportional to $\Phi_{SSC}(\epsilon_{\gamma})$, 
$E^{-\alpha+3}_{\gamma}$ and an exponential
decay term as shown in Eq. (\ref{modifiedsed}).  This
implies that the Fermi accelerated protons interact with the background
photons (in the low energy tail)  of the SSC spectrum. During the April
2004 flaring of Mrk 421, we have shown that the flux from the synchrotron
emission from the high energy $e^+$ and $e^-$ is suppressed
relative to the normal flux implying that the flaring was orphan in
nature, like the one observed in 1ES 1959+650.
We have also shown what type of blazar spectrum will result in orphan flaring and
predict that Mrk 501 and PG 1553+113 are possible candidates for
orphan flaring in the future. 

{Monitoring of these objects by
the TeV gamma-ray telescopes will shed more light on the details on 
the orphan flaring mechanism.


The  work of S.S. is partially
supported by DGAPA-UNAM (Mexico) Project
No. IN110815.

\end{document}